# Few-electron quantum dots for quantum computing

I.H. Chan[a], P. Fallahi[b], A. Vidan[b], R.M. Westervelt[a,b], M. Hanson[c], and A.C. Gossard[c].

[a] Department of Physics, Harvard University, Cambridge, MA 02138, USA
[b] Division of Engineering and Applied Sciences, Harvard University, Cambridge, MA 02138, USA
[c] Materials Department, University of California at Santa Barbara, CA 93106, USA

Two tunnel-coupled few-electron quantum dots were fabricated in a GaAs/AlGaAs quantum well. The absolute number of electrons in each dot could be determined from finite bias Coulomb blockade measurements and gate voltage scans of the dots, and allows the number of electrons to be controlled down to zero. The Zeeman energy of several electronic states in one of the dots was measured with an in-plane magnetic field, and the *g*-factor of the states was found to be no different than that of electrons in bulk GaAs. Tunnel-coupling between dots is demonstrated, and the tunneling strength was estimated from the peak splitting of the Coulomb blockade peaks of the double dot.

## 1. Introduction

Interest in quantum computation has soared since the discovery that quantum computers perform certain calculations exponentially faster than a classical computer (Shor, 1997; Grover, 1997). Demonstrations that quantum computation is experimentally feasible using nuclear magnetic resonance techniques provided experimental motivation (Chuang *et al.*, 1998; Vandersypen *et al.*, 2001), but a more scalable system is desired for quantum computation. Numerous proposals for the physical realization of a quantum computer provide several promising avenues of research in the implementation of quantum computers. These include using the spins of electrons confined in quantum dots (Loss and DiVincenzo, 1998), the nuclear spins of phosphorus atoms embedded in crystalline silicon (Kane, 1998), the electronic states of ions in optical traps (Cirac and Zoller, 1999), and persistent currents in superconducting rings (Mooij *et al.*, 1999). These proposals are challenging experimentally. Recent progress in semiconductor quantum dots includes measurements of individual few-electron quantum dots (Ashoori, 1996; Tarucha *et al.*, 1996; Gould *et al.*, 1999) and double few-electron quantum dots (Elzerman *et al.*, 2003). In addition, the Zeeman energy was measured for electrons in small quantum dots (Hanson *et al.*, 2003; Potok *et al.*, 2003).

In this paper, we discuss measurements on tunnel-coupled few-electron quantum dots, which serve as the building blocks for a quantum computer, as proposed by Loss and DiVincenzo (1998). The quantum bit (qubit) of information would be encoded in the spin of an individual electron that is physically contained inside a small semiconductor quantum dot. Quantum computation would then be performed on an array of quantum dots by sequences of spin-flip and "square-root of swap" (swap$^{1/2}$) operations on the spins. These operations, which can be classified into single-qubit (spin-flip) and two-qubit (swap$^{1/2}$) operations, were shown to be the minimal set needed to construct a universal quantum computer (DiVincenzo, 1995). The spin-flip operation would be carried out using electron-spin resonance (ESR) to flip the spin of individual electrons, while the swap$^{1/2}$ operation would entangle a pair of spins by controlling the tunneling between the quantum dots. Spin-selective filters (Folk *et al.*, 2003) may then be used to read out the results of the calculation. One key experimental hurdle that must be overcome is storing just one electron in each quantum dot. We present data that demonstrates the possibility of storing a single electron in each of the two few-electron dots. The *g*-factor of electron spins was measured and found to be no different from that in bulk GaAs, and tunnel-coupling between two few-electron dots was demonstrated.

## 2. Experimental Setup

The scanning electron micrograph in Figure 1(a) shows a few-electron double dot device used in these experiments. The light gray areas are tunable Cr:Au gates fabricated using electron-beam lithography on a GaAs/AlGaAs heterostructure containing a two-dimensional electron gas (2DEG) located 52 nm beneath the surface. The wafer was grown by molecular beam epitaxy, and consists of the following layers (from the surface down): 50 Å GaAs, 250 Å Al$_{0.3}$Ga$_{0.7}$As, Si δ-doping, 220 Å Al$_{0.3}$Ga$_{0.7}$As, 200 Å GaAs, 1000 Å Al$_{0.3}$Ga$_{0.7}$As, a 20-period GaAs/ Al$_{0.3}$Ga$_{0.7}$As superlattice, a 3000 Å GaAs buffer, and finally the semi-insulating GaAs substrate. This configuration places the 2DEG in a square-well potential. The 2DEG was measured to have a mobility $\mu = 470,000$ cm$^2$/Vs and a density $N = 3.8 \times 10^{11}$ cm$^{-2}$ at 4 K. A square-well was used to get better confinement of the electrons in the *z*-dimension. This allows greater control of the few-electron quantum dots because we are able to electrostatically push the electrons out of the dot using the surface metal gates without leaking electrons out in the *z*-direction. The lithographic size of the dot is nominally $250 \times 260$ nm$^2$.

The few-electron dots were designed such that the tunnel-barriers to the leads were spaced as closely as possible, following the work of Gould *et al.* (1999). This was done to maximize the probability that both leads are connected to the same puddle of electrons inside the dot, as the electrostatic potential in small quantum dots is not perfectly smooth. The mental picture that helps explain why this is so is that of a pond. When a pond (dot) is filled with water (electrons), the pond is a contiguous mass of water. However, when the pond is drained and close to being empty, isolated puddles of water start to form within the pond due to the unevenness of the bottom of the pond (roughness of the potential inside the quantum dot). Having electrical leads close to the same part of a dot therefore increases the chance that both are connected to the same puddle of electrons.

The few-electron dot device was measured in an Oxford Instruments Kelvinox 100 dilution refrigerator with a base temperature $T = 70$ mK ($k_B T = 6$ μeV). The measurement setup is shown in Figure 1(b). The two few-electron dots were in series, so one lock-in amplifier and current preamplifier could be used to measure the conductance of either dot or both dots. Gate voltages were computer controlled. A superconducting magnet inside the Kelvinox 100 Dewar provided an in-plane magnetic field up to 7 T.

## 3. Transport Measurements

We perform a series of experiments to demonstrate that each quantum dot in our few-electron double dot device can contain either zero, one or two electrons, as required by the Loss- DiVincenzo proposal for a semiconductor quantum computer. In our experiments, we first completely emptied the dots of all electrons and then used the Coulomb blockade effect to effectively fill the dots back up, one electron at a time, to the desired number.

## 3.1 Single-Dot Finite Bias Coulomb Blockade Measurements

The number of electrons on each quantum dot can be determined from finite bias Coulomb blockade measurements. When the Coulomb blockade diamond is observed not to close, it indicates that there are no more electrons left in the dot. By energizing only the gates needed to form either dot, the finite bias Coulomb blockade for each dot may be measured separately, as shown in Figure 2. Figure 2(a) shows the differential conductance $dG_{D1}/dV_{DS}$ of Dot 1 as a function of its sidegate voltage $V_{G1}$ and the dc source-drain bias $V_{DS}$, where $G_{D1}$ is the conductance of Dot 1. Figure 2(b) shows the corresponding differential conductance $dG_{D2}/dV_{DS}$ of Dot 2 as a function of its sidegate voltage $V_{G2}$ and the dc source-drain bias $V_{DS}$, where $G_{D2}$ is the conductance of Dot 2. The Coulomb blockade diamonds for both dots are visible as black diamonds, and the additional lines parallel to the sides of the diamonds are Coulomb blockade peaks from excited states of the dots. As the sidegate voltages of both dots are made more negative, the last Coulomb blockade diamond for each dot did not close. No evidence of other Coulomb blockade peaks was seen at drain-source biases out to 9 mV for Dot 1 and 7 mV for Dot 2.

With the dots known to be empty, the absolute number of electrons in each dot can be labeled as indicated by the colored numbers in Figure 2. This therefore enables a single electron to be stored in each dot. Furthermore, Dot 1 can hold up to five electrons while Dot 2 can hold up to three electrons, indicating that these dots are very small. The small number of electrons on each dot is easily explained by noting that the electrostatic potential from the gates that define a small dot is shallower than that of a larger dot, in addition to being smaller in the lateral dimensions. Indeed, the dimensions of few-electron quantum dots are larger than anticipated and therefore easier to fabricate.

## 3.2 Three-Dimensional Scan of a Few-Electron Dot

A drawback of using the finite bias Coulomb blockade technique for laterally-defined dots is that stray capacitance couples the sidegate voltage to the tunnel barriers of the dot and causes them to get progressively more opaque. This problem of a diminishing conductance signal can be overcome by re-tuning the quantum point contact (QPC) voltages, thereby controlling the tunnel barrier heights, every time the sidegate voltage is changed. The resulting three-dimensional scan in the sidegate and QPC voltages, which are all the gate voltages that may be adjusted, also shows that the Coulomb blockade peaks end and therefore that our dots can be emptied of electrons.

Figure 3 shows the three-dimensional conductance data of Dot 2 as a series of two-dimensional scans in QPC gate voltages $V_{QPC1}$ and $V_{QPC2}$ as sidegate voltage $V_{G2}$ goes from (a) –0.500 V to (f) –1.625 V in steps of –0.225 V. Each two-dimensional scan is a re-tuning of the QPC gate voltages to Dot 2, and maximizes the Coulomb blockade peak height of the Coulomb blockade peaks within the scan.

As sidegate voltage $V_{G2}$ is made more negative in Figures 3 (a) to (f), the Coulomb blockade peaks are shifted toward the top right as it progressively takes less QPC voltage to induce the same number of electrons onto Dot 2. However, no new Coulomb blockade peaks are revealed in the sequence of figures, leading to the conclusion that Dot 2 is empty past the last Coulomb blockade peak (toward the bottom left corner of each figure). The number of electrons in Dot 2 may therefore be labeled as indicated, and shows that Dot 2 can contain up to three electrons before it merges with its leads. By choosing an appropriate combination of QPC and sidegate voltages, the number of electrons in Dot 2 may therefore be selected.

## 4. Zeeman Energy of Few-Electron States

It is also important to learn about the properties of electron spins for implementing the spin-flip operation with ESR. Given the isolated nature of electrons in few-electron quantum dots compared with electrons in bulk semiconductor, it is not obvious that the spin of these electrons will behave like those of electrons in a bulk semiconductor. In order to determine the *g*-factor of electrons inside a few-electron dot, the Zeeman energy of several electronic states was measured.

The Zeeman energy of an electron may be measured as a component of the electron's addition energy. The addition energy $E_A$ of electrons in an in-plane magnetic field is the sum of the charging energy $E_C$, the level spacing $\boldsymbol{D}e$, and the Zeeman energy $E_Z$:

$$E_A = E_C + \boldsymbol{D}e + E_Z, \qquad (1)$$

where $E_Z = \pm g\mu B_\parallel$, $g$ is the *g*-factor, $\mu$ is the Bohr magneton, and $B_\parallel$ is the in-plane magnetic field. The in-plane magnetic field should not affect the orbital motion of electrons in the 2DEG because the square-well potential is narrow (20 nm wide) in the *z*-dimension. Therefore, the charging energy $E_C$ and level spacing $\boldsymbol{D}e$ should not change with in-plane magnetic field. The only component of the addition energy that depends on magnetic field is the Zeeman energy $E_Z$. This means that the sidegate voltage separating Coulomb blockade peaks $\boldsymbol{D}V_{G2}$ (the peak spacing) is related to the Zeeman energy by the relation:

$$E_A = e\boldsymbol{D}V_{G2}\, C_G/C_\Sigma$$
$$= \pm g\mu B_\parallel + \text{constant in } B_\parallel, \qquad (2)$$

where $C_G$ is the sidegate-to-dot capacitance and $C_\Sigma$ is the total dot self-capacitance. One can therefore measure the conductance of a dot as a function of in-plane magnetic field and sidegate voltage to obtain the Zeeman energy, and thus the *g*-factor.

Coulomb blockade conductance measurements used to determine the Zeeman energy are shown in Figure 4(a), where the conductance $G_{D2}$ of Dot 2 was measured as a function of in-plane magnetic field $B_\parallel$ and sidegate voltage $V_{G2}$. Four Coulomb blockade peaks are clearly visible in the figure; a fifth peak is very low in amplitude, and no useable data was extracted from it. The positions of these four peaks in sidegate voltage $V_{G2}$ were computed by curve-fitting each peak to the expected conductance lineshape, as given by Beenakker (1991) and Meir *et al.* (1991). This is shown in Figure 4(b). The peak spacings $\boldsymbol{D}V_{G2}$ between the Coulomb blockade peaks are shown in Figure 4(c), and are observed to slope with magnetic field. The Zeeman energy traces for $|g| = 0.44$ (expected for electrons in bulk GaAs) are shown by the dashed lines, and were computed via Equation (2). A "lever-arm" ratio of capacitances $C_G/C_\Sigma = 1/90$ was used in the calculation, which was obtained from the ratio of the half-height to the full-width of the Coulomb blockade diamond of Dot 2 [Figure 2(b)]. Figure 4(c) shows that the measured and the expected Zeeman energies are in agreement, and therefore that the *g*-factor of electrons in a few-electron dot is consistent with the *g*-factor $|g| = 0.44$ of electrons in bulk GaAs.

Recent data from Hanson *et al.* (2003) also finds that the *g*-factor in few-electron dots is similar to that in bulk GaAs at fields below about 7 T, but indicates that there is a saturation in the Zeeman energy at higher fields (up to the measured 15 T). Further work needs to be done to clarify the origin of this behavior.

## 5. Tunnel-Coupled Few-Electron Dots

An important aspect of the Loss-DiVincenzo proposal (Loss and DiVincenzo, 1998) is the use of tunnel coupling between quantum dots to entangle electron spins (swap$^{1/2}$ operation). In this operation, both the strength and the duration of the coupling must be controlled. In this section, we demonstrate that tunnel-coupling between dots can be achieved, and deduce the strength of the tunneling from the Coulomb blockade peak splitting.

Figure 5 shows the conductance $G_{DD}$ of Dot 1 and Dot 2 in series as a function of the two sidegate voltages $V_{G1}$ and $V_{G2}$. The Coulomb blockade peaks of the two dots in series are clearly observed to be split, and the hexagonal charge-stability pattern characteristic of coupled double dots is highlighted as shown. By measuring the fractional peak splitting $f \equiv \Delta V_S/\Delta V_P$, the actual tunnel coupling between the dots in Figure 5 can be estimated. Following Livermore *et al.* (1996) and interpolating between the two theoretical formulas appropriate for weak and strong tunnel-coupling (Golden and Halperin, 1996a and 1996b), an inter-dot conductance of 1.2 $e^2/h$ is estimated from the measured fractional peak splitting $f = 0.3$ from Figure 5.

In making this estimate, the capacitive-coupling between the few-electron dots was assumed to be negligible. A finite capacitive-coupling would alter the scaling of $f$ to give (Golden and Halperin, 1996a):

$$(1 - f_C) = \frac{C_\Sigma}{C_\Sigma + 2C_{INT}}(1 - f) \qquad (3)$$

where $f$ is the fractional peak splitting given by theory, which does not include capacitive coupling, and $f_C$ is the fractional peak splitting that includes the effect of inter-dot capacitance $C_{INT}$. The inter-dot capacitance $C_{INT}$ may be found by setting the inter-dot tunnel-coupling as low as possible, and calculating $C_{INT}$ using (Chan, 2003)

$$\frac{C_{INT}}{C_\Sigma} = \frac{f}{2 - f} \qquad (4)$$

from the peak splitting that remains.

## 6. Conclusions

The state of implementing a complete Loss-DiVincenzo quantum computer is still in its infancy, but the first experimental steps have been taken. Quantum dots containing just a few electrons have been fabricated, and the number of electrons in these dots can be determined and controlled. The Zeeman energy of electrons in these few-electron dots indicate that the behavior of electron spin in such small dots is no different from that in bulk GaAs (at least up to 7 T). Tunnel-coupling of few-electron dots, essential for implementing the swap$^{1/2}$ operation, was demonstrated, and the strength of the coupling was estimated from the fractional peak splitting.

The next steps in researching semiconductor quantum computers will involve electron-spin resonance (ESR) measurements, measuring spin coherence lifetimes, and controlling both the strength and the duration of tunnel-coupling between dots. Controlling the tunnel-coupling is probably the most challenging and interesting task facing semiconductor quantum computers, as it goes to the heart of quantum computation, which is entanglement. The activity in quantum computing research will intensify, and many exciting results will develop in the coming years.

**Acknowledgements**
We thank L. Kouwenhoven, D. Loss, C. Marcus, and S. Tarucha for helpful discussions. This work was supported by DARPA QuIST program (DAAD19-01-1-0659) at Harvard, iQUEST at UC Santa Barbara, the Nanoscale Science and Engineering Center (NSF PHY-01-17795) based at Harvard, and by the Center for Imaging and Mesoscale Structures (CIMS) at Harvard.

# References

Ashoori, R.C. (1996). *Nature* **379**, 413.

Beenakker, C.W.J. (1991). *Phys. Rev. B* **44**, 1646.

Chan, I.H. (2003) *Quantum Dot Circuits: Single-Electron Switch and Few-Electron Quantum Dots*, Ph.D. Thesis, Harvard University.

Chuang, I.L., L.M.K. Vandersypen, X. Zhou, D.W. Leung, and S. Lloyd (1998). *Nature* **393**, 143.

Cirac, J.I. and P. Zoller (1999). *Nature* **404**, 579.

DiVincenzo, D.P. (1995). *Phys. Rev. A* **51**, 1015.

Elzerman, J.M., R. Hanson, J.S. Greidanus, L.H.W. van Beveren, S. de Franceschi, L.M.K. Vandersypen, S. Tarucha, and L.P. Kouwenhoven (2003). *Phys. Rev. B* **67**, 161308.

Folk, J.A., R.M. Potok, C.M. Marcus, and V. Umansky (2003). *Science* **299**, 679.

Golden, J.M. and B.I. Halperin (1996a). *Phys. Rev. B* **53**, 3893.

Golden, J.M. and B.I. Halperin (1996b). *Phys. Rev. B* **54**, 16757.

Gould, C., A.S. Sachrajda, P. Hawrylak, P. Zawadzki, Y. Feng, and Z. Wasilewski (1999). *Proceedings of the Fifth International Symposium on Quantum Confinement: Nanostructures*, 270.

Grover, L.K. (1997). *Phys. Rev. Lett*. **79**, 325.

Hanson, R., B. Witkamp, L.M.K. Vandersypen, L.H.W. van Beveren, J.M. Elzerman, and L.P. Kouwenhoven (2003). cond-mat/0303139.

Kane, B.E. (1998). *Nature* **393**, 133.

Livermore, C., C.H. Crouch, R.M. Westervelt, K.L. Campman, and A.C. Gossard (1996). *Science* **274**, 1332.

Loss, D., and D.P. DiVincenzo (1998). *Phys. Rev. A* **57**, 120.

Meir, Y., N.S. Wingreen, and P.A. Lee (1991). *Phys. Rev. Lett*. **66**, 3048.

Mooij, J.E., T.P. Orlando, L. Levitov, L. Tian, C.H. van der Wal, and S. Lloyd (1999). *Science* **285**, 1036.

Potok, R.M., J.A. Folk, C.M. Marcus, V. Umansky, M. Hanson, and A.C. Gossard (2003). *Phys. Rev. Lett.* **91**, 016802/1

Shor, P. (1997). *SIAM J. Comput*. **26**, 1484.

Tarucha, S., D.G. Austing, T. Honda, R.J. van der Hage, and L.P. Kouwenhoven (1996). *Phys. Rev. Lett.* **77**, 3613.

Vandersypen, L.M.K., M. Steffan, G. Breyta, C.S. Yannoni, M.H. Sherwood, and I.L. Chuang (2001). *Nature* **414**, 883.

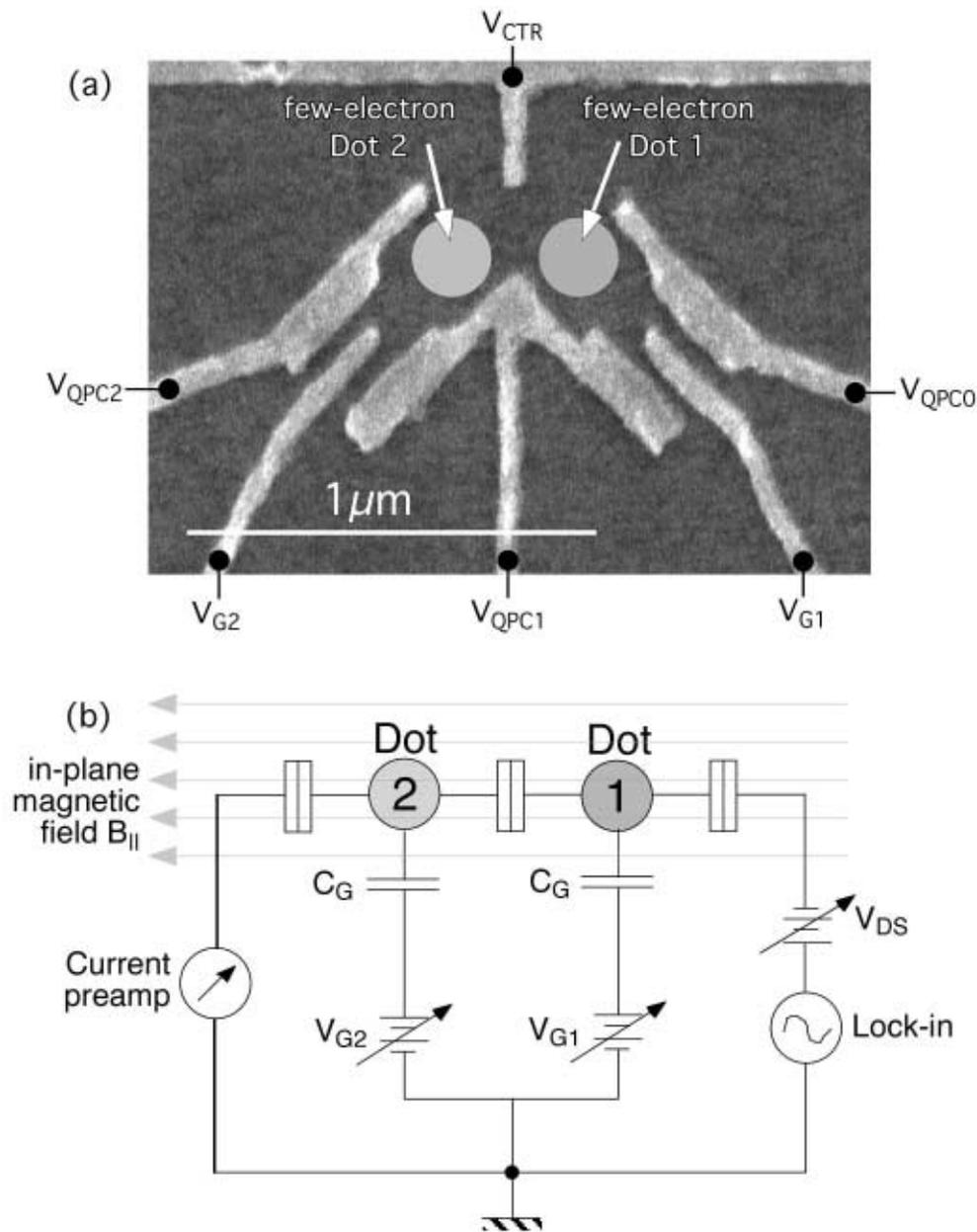

**Figure 1** (a) SEM photograph of the few-electron double dot device. The light areas are metal gates used to define the quantum dots. The locations of the dots are highlighted by circles - red for Dot 1 and green for Dot 2. The gate voltage applied to each gate is labeled by the black text. (b) Measurement setup for the few-electron quantum dots. The conductance of Dot 1, Dot 2, or both in series was measured using conventional lock-in techniques. A drain-source DC bias voltage could be applied across the dot(s) for finite bias Coulomb blockade measurements. An in-plane magnetic field of up to 7 T could also be applied. Not shown are voltages for controlling tunnel barrier heights of the dots.

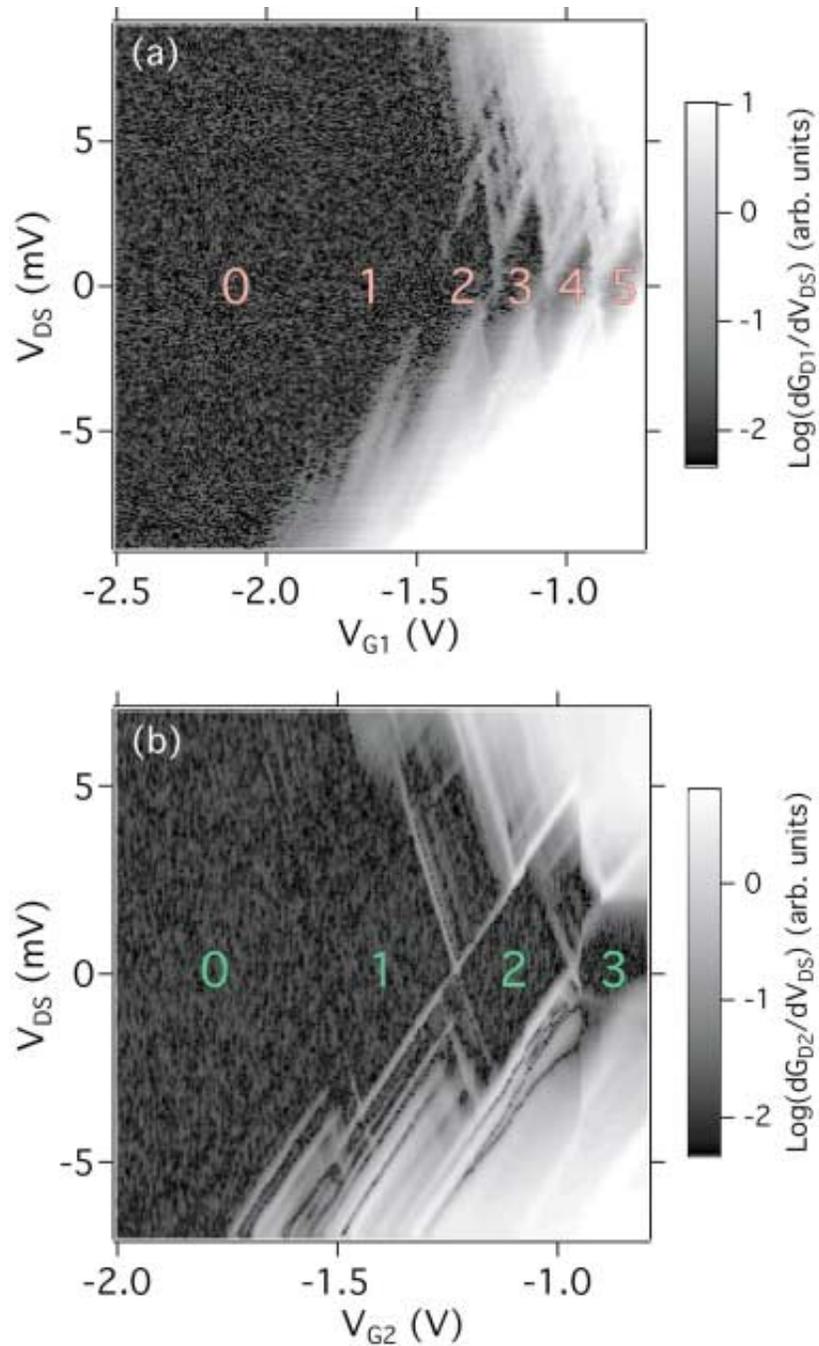

**Figure 2** Finite bias Coulomb blockade measurements of (a) Dot 1 and (b) Dot 2, acquired separately. The left-most Coulomb blockade diamonds do not close, indicating that no more electrons are to be found inside the dots. This allows the number of electrons in each dot to be labeled as indicated. The dashed lines are extrapolations of Coulomb blockade peaks that are visible at high drain-source voltages.

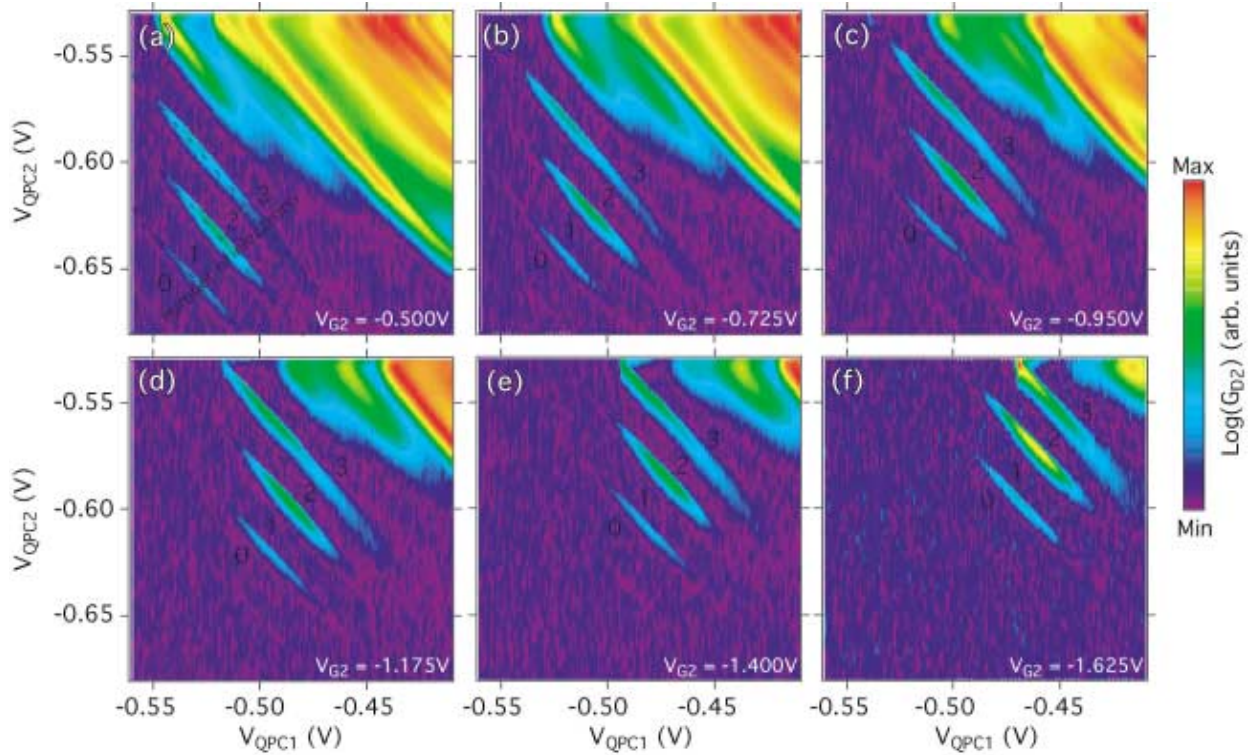

**Figure 3** Conductance $G_{D2}$ of Dot 2 as a function of QPC voltages $V_{QPC1}$ and $V_{QPC2}$, and sidegate voltage $V_{G2}$. Each scan (a) to (f) in QPC voltages is a re-tuning of the tunnel barriers to Dot 2 at a different sidegate voltage. This re-tuning maximizes the Coulomb blockade peak heights for greatest peak visibility. As sidegate voltage $V_{G2}$ is decreased from (a) -0.500 V to (f) -1.625 V, the Coulomb blockade peaks are shifted toward the top right. However, no new Coulomb blockade peaks are revealed from the bottom left, indicating that there are no more electrons left in the dot. The number of electrons in Dot 2 may thus be labeled as indicated, and the number of electrons in the dot may be

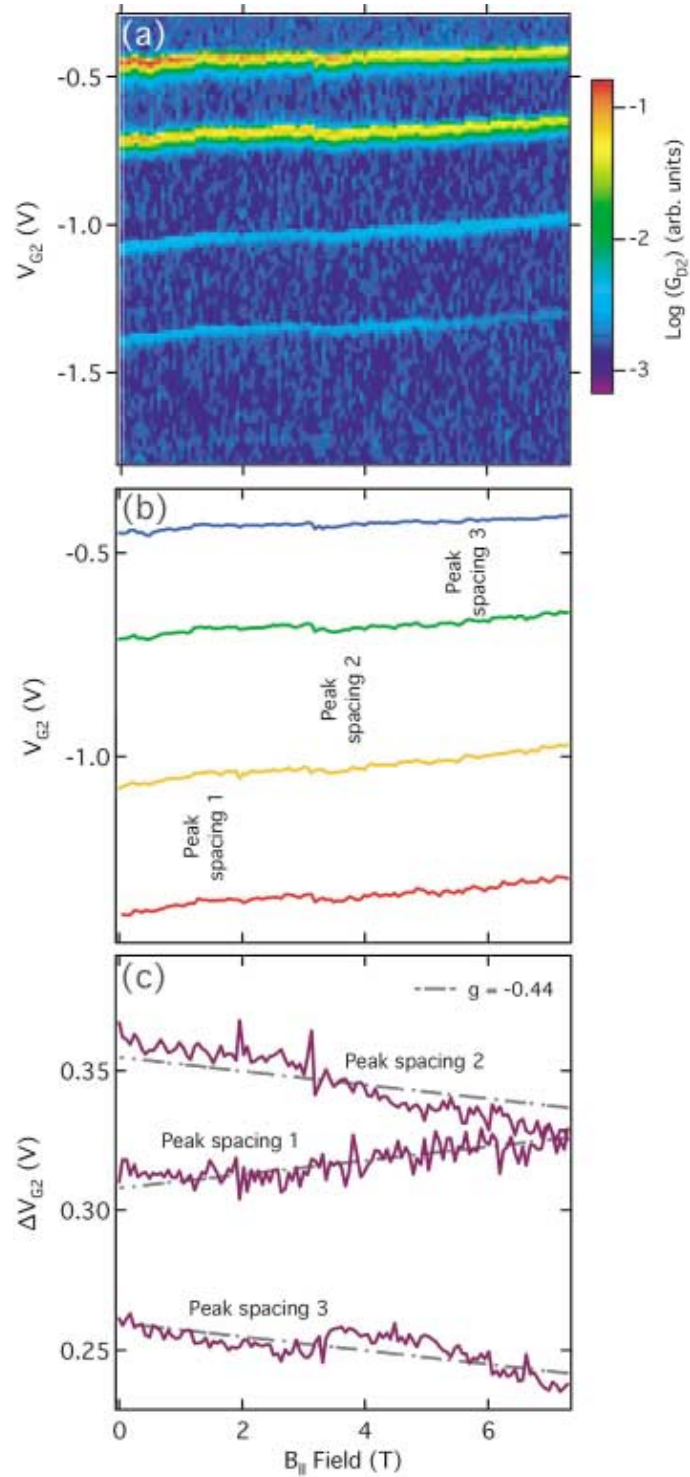

**Figure 4** Extraction of Zeeman energy data from Coulomb blockade peak spacings. (a) Conductance of Dot 2 as a function of in-plane magnetic field and sidegate voltage. (b) Peak location of the four Coulomb blockade peaks, which was obtained by curve-fitting the data in (a). (c) Peak spacings of the Coulomb blockade peaks as a function of in-plane magnetic field. The Zeeman energy in bulk GaAs (dashed lines) assume a g-factor |g| = 0.44, and were calculated using Equation (2).

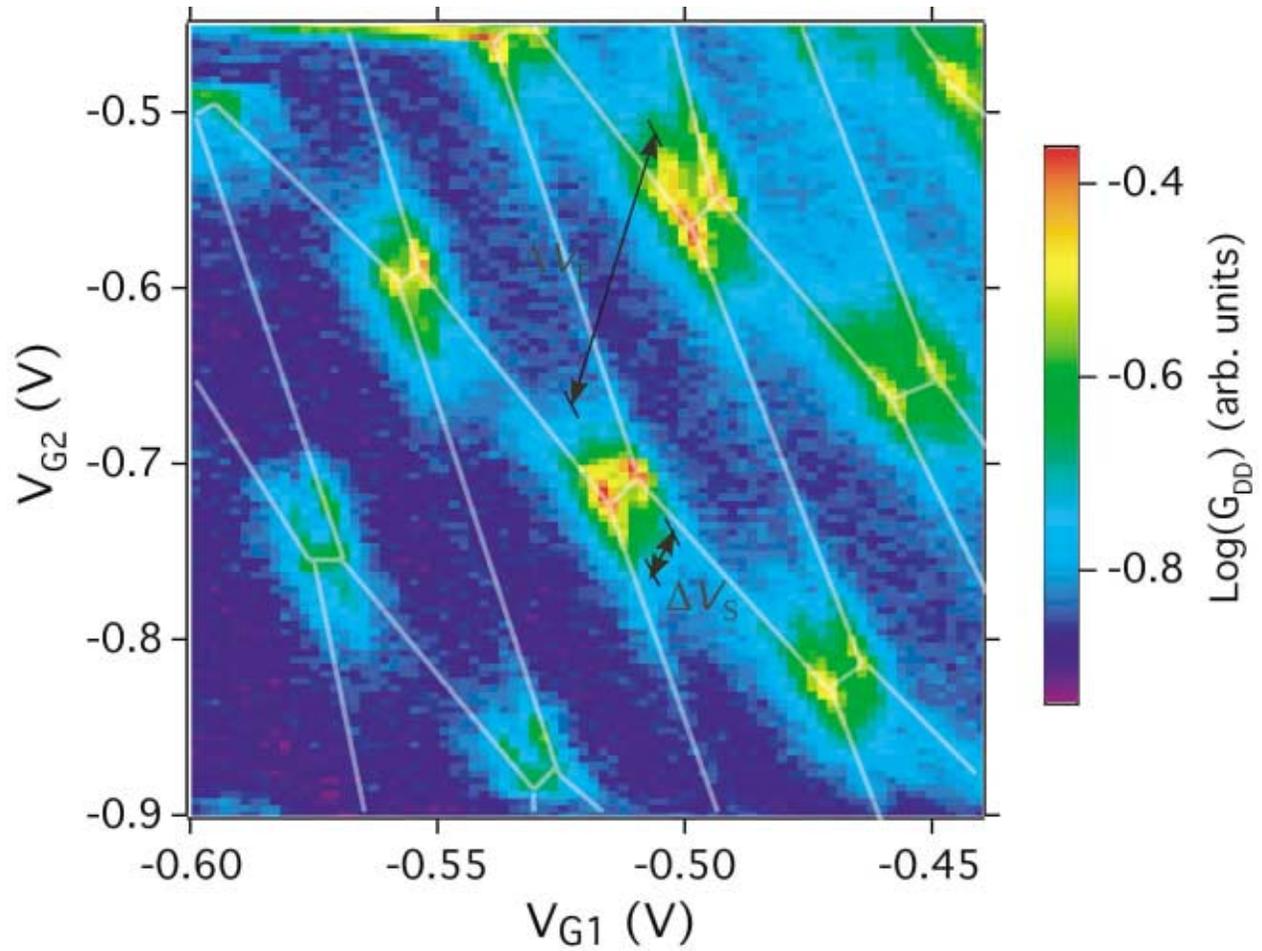

**Figure 5** Conductance of tunnel-coupled few-electron dots in series as a function of sidegate voltages. The Coulomb blockade peaks are clearly split, and the hexagonal charge-stability pattern is highlighted. This hexagonal pattern is indicative of coupled double dots, and a fractional peak splitting $f = 0.3$ was measured.